# H⁻ Beam formation simulation in negative ion source for CERN's Linac4 accelerator

**A. Vnuchenko**[a,1], **J. Lettry**[a], **S. Mochalskyy**[b], **D. Wünderlich**[b], **U. Fantz**[c], **M. Lindqvist**[b], **A. Revel**[c] **and T. Minea**[c]

[a] *European Organization for Nuclear Research (CERN)*
   *Geneva, Switzerland*
[b] *Max-Planck Institut für Plasmaphysik*
   *Address, Garching, Germany*
[c] *Laboratory of Physics of Gases and Plasma (LPGP), CNRS, Université Paris-Saclay*
   *Orsay, France*
   *E-mail:* anna.vnuchenko@cern.ch

ABSTRACT: The caesiated surface negative ion (H⁻) source is the first element of CERN's LINAC4 a linear injector designed to accelerate negative hydrogen ions to 160 MeV. The IS03 ion source is operated at 35 mA beam intensity and reliably feeds CERN's accelerator chain, H⁻ ions are generated via plasma volume and caesiated molybdenum (Cs-Mo) plasma electrode surface mechanisms. Studying the beam extraction region of this H⁻ ion source is essential for optimizing the H⁻ production. The 3D Particle-in-cell (PIC) Monte Carlo (MC) code ONIX (Orsay Negative Ion eXtraction [1]), written to study H⁻ beam formation processes in neutral injectors for fusion, has been adapted to single aperture accelerator H⁻ sources. The code was modified to match the conditions of the beam formation and extraction regions of the Linac4 H⁻ source [2]. A set of parameters was chosen to characterize the plasma and to match the specific volume and surface production modes. Simulated results of the extraction regions are presented and benchmarked with experimental results obtained at the Linac4 test stand [3].

---

[1] Corresponding author.

# Contents

## 1. Introduction

Linac4 operates with negatively charged hydrogen ions beams produced by a Radio Frequency Inductively Coupled Plasma (RF-ICP) type ion source, composed of a ceramic plasma chamber surrounded by an external five-turn RF coil. The beam is extracted by a five-electrodes extraction system. A puller-dump electrode operated at 2-3 kV/mm relative to the source, extracts the H$^-$ beam and electrons. The filter field reduces the energy of electrons present in the beam formation region upstream of the extraction apertures, where low energy electrons contribute to the dissociative attachment process. The dump field deflects the extracted electrons, and the H$^-$ beam is then accelerated to the 45 keV energy. The filter and dump fields are generated by pairs of permanent magnets located around the Plasma Electrode (PE) and in the puller dump electrode. A detailed information of the IS03b ion source and operation are given in [2, 4].

The ion source delivering this H$^-$ beam is based on two mechanisms: the "volume" (dissociative attachment of a low energy electron to an excited $H_2^v$ molecule) and "plasma surface" (re-emission as H$^-$ ion of a proton or hydrogen atom produced in the bulk plasma and impacting onto a low work function Cs coated Mo-PE surface) [5-7]. The ability of Cs to release electrons towards the impinging hydrogen atom depends on the coverage fraction of the surface. The Cs layer on the plasma facing surface of the PE reduces the work function from about 4.3 eV for pure Mo surface to a value even below the 2.2 eV corresponding to metallic Cs. Depending on the mode of operation, the ratio between electrons and ions (e/H$^-$) is in the range of 20 to 30 in volume mode and this ratio can be reduced to around one in surface mode.

The beam formed in this source results from the convolution of volume and surface ion production. The beam formation starts in the vicinity to the PE of the ion source, where charge separation appears forming a negative sheath. The puller electrode sets an electric field that extracts H$^-$ and electrons simultaneously and repels positively charged particles back towards the expansion chamber to build the so-called meniscus. The shape of the meniscus and the ions produced on the PE-surface closest to PE-aperture determine the initial properties of the beam to be propagated along beam optical components.

The goal of the numerical simulations and associated experimental program is to gain insight into the initial beam properties resulting from these different ion production modes. The ONIX PIC- Monte Carlo collision (MCC) software is used to simulate the particle trajectories from the plasma toward the extraction region and the extracted beam. 3D models are applied due to impact of magnetic fields close to extraction region, breaking any spatial symmetry in the system.

## 2. Simulation model

3D PIC-MCC code ONIX is used to simulate the hydrogen plasma and the extracted particle features in the vicinity of the PE. The code has been previously validated and applied to simulate the extraction of negative ions [8]. The code is self-consistent and parallelized via the message passing interface (MPI) library [9] using a domain particle decomposition [10]. Charged particles are presented by macro-particles and their charge is interpolated onto a regular mesh.

This code, originally dedicated to ITER's neutral injector sources, has been modified to match single aperture sources. New non-periodic boundary conditions of the simulation volume in directions orthogonal to the beam axis and geometry of the standard IS03 of CERN's Linac4 H⁻ source have been implemented. Plasma particles striking these boundaries are reinjected into the bulk plasma. In addition, particles hitting the left (x=0) boundary are mirrored into the calculation domain. The extraction potential is applied to the right boundary of the simulation domain in a plane orthogonal to the beam axis. The remaining domain boundaries potential are set to zero. Model ONIX is applied to the beam formation region in direct vicinity of the PE since a complete 3D modelling of the full ion source under realistic plasma parameters is beyond today's computer capabilities. The geometry of the computational domain corresponding to IS03 is illustrated in Figure 1.

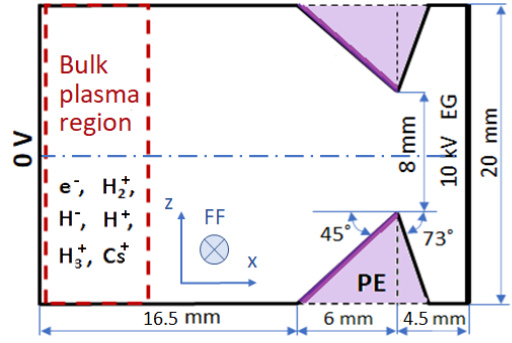

**Figure 1.** Schematic view of the simulation domains used in the ONIX code (x–z mid plane) for modelling beam formation of IS03b standard PE. The filter field (FF) orientation is indicated. The bore diameter of PE is 8 mm.

At the beginning of the simulation, the particles are injected in the so-called "bulk plasma region". Particles inside this region have a Maxwellian energy distribution and the particles outflow is constant over time. The ONIX code uses realistic plasma and source conditions. The plasma densities present in the H⁻ sources close to the PE is about $10^{17}$ m$^{-3}$. The set of plasma parameters is taken from the numerical simulations obtained for the Linac4 IS03 [11]. The H⁻ current is highly dependent on the caesiation of the sources that determines the work function of the PE-wall and, accordingly, the surface production rate. The external applied 3D magnetic field topology [12] and the extraction potential at the right-hand side border of the simulation domain are also specified. The contribution from the volume mode to the extracted current directly depends on the negative ion density in the bulk plasma region.

The code uses macro-particles representing $5\times10^4$ real particles to reduce real computation time. The spatial limitation of the simulation domain is calculated to take into account the space charge of each macro particle. The charge of the macro particles of the simulation domain is linearly interpolated and distributed onto the PIC nodes. The computational domain is discretized into a regular grid of 416×312×312 cells with cell size of 6.5×10⁻⁵ m, slightly larger the Debye length ($\lambda_D \approx 4.1 \times 10^{-5}$ m) to avoid nonphysical numerical increase of kinetic energy of the charged particles observed in simulations until the effective Debye length is of the same order as the grid size. The chosen time step is $5 \times 10^{-12}$ s. that should be smaller than the inverse plasma frequency of $3.41 \times 10^{-11}$ s [13]. The numerical parameters are chosen to reduce the required CPU time without losing numerical accuracy.

The plasma initially expands into the 'empty' simulation domain where the extraction field freely penetrates; the populations of the plasma migrate according to their mass dependent velocities. The beam formation plasma then stabilizes in the vicinity of the PE aperture and forms the self-consistent meniscus. The meniscus position and its curvature define the initial radial velocity and angle of the trajectory of each particle extracted from the beam formation region. Previous studies have shown that the meniscus shape and position depend on applied extraction potential, geometry of the PE, plasma bulk density and H⁻ emission rate [8, 14]. The particle is considered extracted when crossing the right boundary of the simulation domain. Rapid increase of electron current is observed at the beginning of the simulation. After a transitory phase due to low electron mass, the system is reaching steady state and stable electron and H⁻ currents.

## 3. Result and discussions

The self-consistent meniscus is formed in the vicinity of the PE aperture. The meniscus position and the depth of its curvature play an important role for the beam formation since it defines the velocity and angle of trajectory of each extracted particle. A fraction of the PE appears to be located between the meniscus and the PE-extraction aperture, H⁻ ions originating from this few mm ring shaped region could possibly be at the origin of a beam halo [15]. The density distribution of positive ions in the axial vertical (x-y) and axial horizontal (x-z) planes of the IS03 simulation domain are shown in Figure 2, after steady state was reached. The vertical magnetic filter field breaks the symmetry along the horizontal plane due to impact on the electron flow, that influence the distribution of the positive charged species. This initial anisotropic distribution will affect the extracted beam.

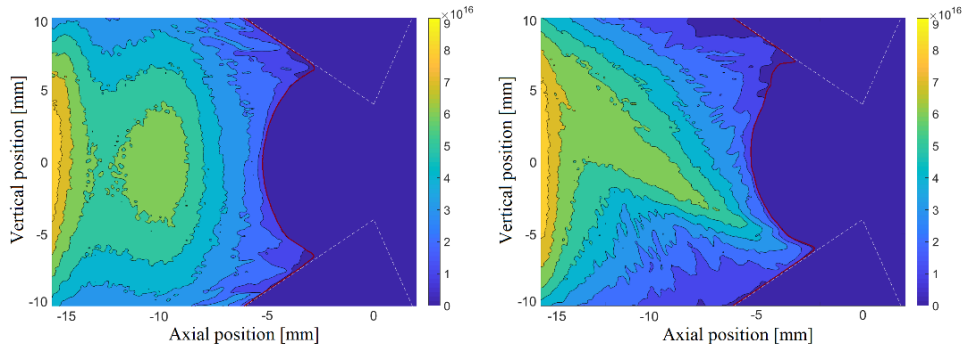

**Figure 2.** Density maps of positively charged particles ($H^+$, $H_2^+$, $H_3^+$) in the vetrical (x-y) (left) and horizontal (x-z) (right) central axial planes. The red curve represents the meniscus shape.

ONIX simulations were performed for various initial parameter sets to investigate their impact on the extracted beams. Figure 3 shows the evolution of the H⁻ and co-extracted electron beam currents during simulation for H⁻ surface emission rates of 20 Am⁻² and 80 Am⁻², and bulk plasma densities of $10^{17}$ m³ in the injection region with the electron to H⁻ ion density ratio (e:H⁻) of e:H⁻ = 1:1. A 10 kV extraction potential is applied. The temperature of electron is 2 eV and H⁻ is 1.5 eV. These parameters were chosen based on computational [11] and experimental [16] data for the current IS configuration.

Surface production of H⁻ is implemented assuming uniform flux from the PE surface in these simulations. The value of currents of 41 and 63.5 mA correspond to the experimental values that can be obtained for IS03 measured at Linac4 test stand. A steady state of the simulation is reached, however, the co-extracted electron current is oscillating in a chaotic manner around its equilibrium value due to turbulent electron transport through the magnetic fields [17]. The e/H⁻ current ratio is about 1, that corresponds to typical values for a well-caesiated source. The H⁻ current includes volume (32.5 and 28.5 mA) and surface production modes (8.5 and 35 mA). The simulation indicates that emission rate defines the H⁻ extracted from surface of the PE but also moderately impacts the volume contribution. In comparison of the presented cases, the difference is about 12 %.

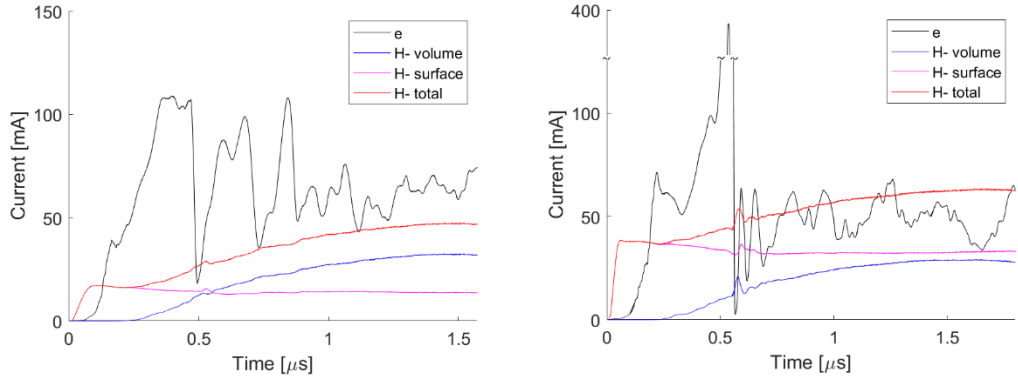

**Figure 3.** The total extracted H⁻ current (red line) extracted from volume (blue line) and plasma surface (pink line) production modes and co-extracted electron (black line) current for the IS03 system over simulation time using a 10 kV extraction voltage and a magnetic field of 11 mT for H⁻ surface emission rates of 20 Am$^{-2}$ (left) and 80 Am$^{-2}$ (right) at 3.5 mm from PE-tip.

A parametric study of H⁻ production at the Cs covered PE surface changing the surface emission rate is performed for equal initial density of H⁻ and electrons, see Figure 4. The simulation was performed for a variation of the H⁻ surface emission rate from 10 to 500 Am$^{-2}$. This distribution is based on the assumptions of a homogeneous neutral flux towards this surface. The plasma parameters are kept constant in order to investigate general physical effects.

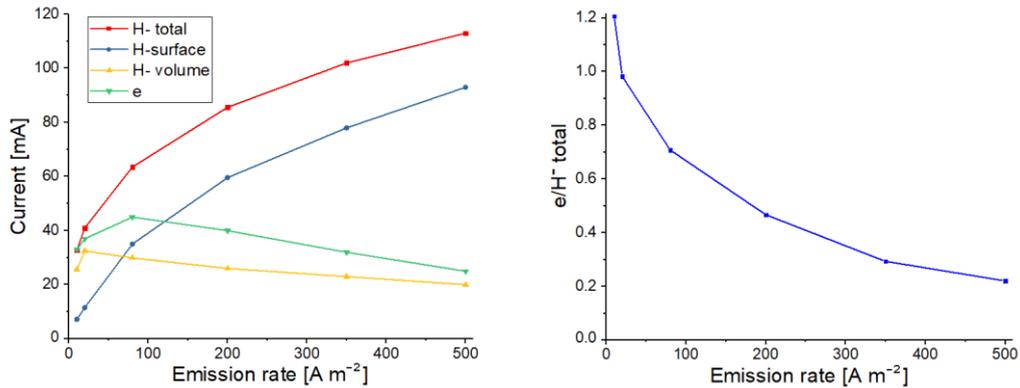

**Figure 4.** Current densities for extracted electrons and H⁻ (left) and e/H⁻ ratio (right) dependence on surface emission rate.

As expected, the emission rate defines the extracted H⁻ originating from the PE surface but, for emission rates above 100 A/m², we also observe a saturation, a minor reduction from volume originating H⁻ ions and of the amount of co-extracted electron. With increasing the negative ion emission rate, the extracted current density continuously grows that leads to significant reduction of e/H⁻ ratio less than 1. Such saturation demonstrate that the plasma requires increased electron energy and density to reach higher surface production rate. Simulation with an emission rate below 10 Am$^{-2}$ is not considered in this work, since in this case volume mode prevails.

The emission of negative ions results in a reduced sheath potential and ultimately to the formation of a potential well. Depending on the depth of this potential well, repels a significant amount of H⁻ back to the plasma where they are destroyed. This leads to a decrease H⁻ from the volume production. The deepness of the potential well relates to the emission rate. Produced H⁻ is directly corelated with plasma density which determines the penetration of the meniscus close to extraction area. However, the H⁻ current is not proportional to surface emission since a significant fraction of the H⁻ are directly extracted through the edges of the meniscus and thus do not affect the charge balance close to the extraction area. Therefore, using a H⁻ emission rate of about 80 Am$^{-2}$, H⁻ are formed with similar contribution of volume and surface components for given plasma parameters.

Figure 5 shows the emission rate dependence on potential well for a line in the vertical-axial plane, 1 mm away from the edge of the calculation domain and an example of axial potential profile for the surface emission rate of 80 Am$^{-2}$. The values of the potential well correspond to a certain probability for surface produced negative ions to reach the plasma volume.

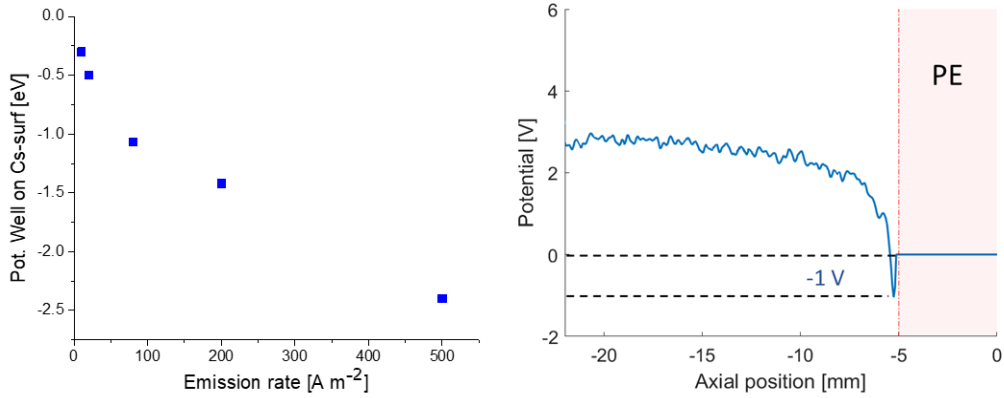

**Figure 5.** Potential well resulting from re-emission of H- ion form the low work function surface and example of potential profile for the surface emission rate of 80 Am$^{-2}$.

The density profiles of H⁻ in the (x-y) and (x-z) planes at the extraction area of simulation domain are shown in Figure 6, after steady state was reached. The asymmetry of the beam profile is revealed, especially in the (x-z) plane for H⁻ production from PE surface caused by the magnetic field. The reason is that the filter field is predominant in the y direction and limits the electron flow, that influences the distribution of the positive charged species in the extraction area.

These asymmetries were estimated and compared with experimental measurement performed at Linac4 test stand [18], see Table 1. The comparison of asymmetry is made using the ratio of the beam contained in π/2 opposite sectors normalized to the average intensity. ONIX simulation with emission rate of 80 mA$^{-2}$ demonstrate asymmetry of ± 5 % of H⁻ beam profile in the horizontal plane and ± 2 % in the vertical plane, while depending on the origin, H⁻ from the surface is about 10 % in the horizontal direction and 2 % in the vertical direction. H⁻ volume produced particles has ± 2-3 % in both orientations. Beam Emission Spectroscopy (BES) measurements show a ± 12% asymmetry in the horizontal plane and ± 1% in the vertical plane.

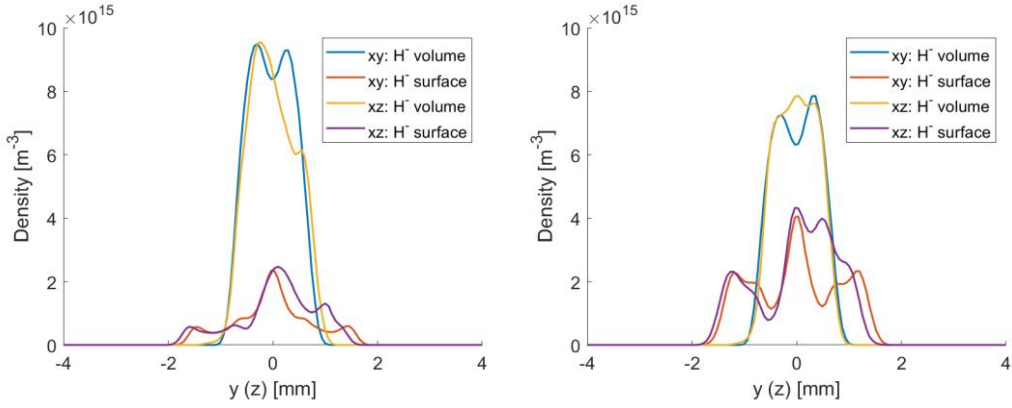

**Figure 6.** Density profile of H⁻ according to their origin (volume and surface production) in the extraction area for H⁻ surface emission rates of 20 Am$^{-2}$ (left) and 80 Am$^{-2}$ (right).

**Table 1.** Asymmetry of beam profiles induced by the filter field: The beam parameters of 20 and 80 A/m$^2$ surface emission rates extracted form ONIX simulations are calculated at 3.5 mm after the PE-tip (beam energy: ≈7-8 keV). The BES measurement is located 37 cm after the PE electrode (beam energy: 45 keV). The beam fractions contained in each of the 4 quadrants (Up, Down, Left, Right) are given.

| Parameters | Units | BES | ONIX at 20 Am$^{-2}$ | | | ONIX at 80 Am$^{-2}$ | | |
|---|---|---|---|---|---|---|---|---|
| | | | Total | Volume | Surface | Total | Volume | Surface |
| **H⁻ current** | mA | **50** | **41** | **33** | **8** | **63.5** | **28.5** | **35** |
| Up | % | 101 | 98 | 98 | 102 | 102 | 103 | 102 |
| Down | % | 99 | 102 | 102 | 98 | 98 | 97 | 98 |
| Left | % | 88 | 101 | 107 | 83 | 95 | 102 | 90 |
| Right | % | 112 | 99 | 93 | 117 | 105 | 98 | 110 |

BES measurements and ONIX simulations demonstrate good agreement between the results. The difference in asymmetry cannot be compared directly since the beam profiles are made at different distances from the plasma electrode. A telescope was installed on two view ports located at 37 cm from the ion source (minimum possible distance from the source) at an angle to beam axis, the light emitted in a 12 mm diameter cylinder is collected onto a fiber and analyzed with a spectrometer [18]. BES measurement reproduces the fraction of the beam profile limited by local telescope capture efficiency. The beam intensity was approximately 50 mA and e/H ≈ 2-3.

The H⁻ yield depends on the surface work function resulting from the surface coverage of Cs on the Mo-PE surface. In simulations, this coverage is specified in terms of surface emission rate which is uniform over the surface and constant over time. However, the recent measurements of a coverage gradient on the Cs-Mo electrode surface taken to determine the distribution of Cs on the PE surface for this source demonstrated that the Cs coating on the surface is not uniform [19]. Therefore, additional simulations to illustrate the impact of variable emission rate distributions along the PE on the total H⁻ production were performed for the cases where the emission rate is constant, with a strong positive (from 2 to 37 A/m$^2$) and negative (from 37 to 2 A/m$^2$) gradient. The average emission rate is 20 A/m$^2$, see Table 2. A plasma density of $10^{16}$ m$^{-3}$ is chosen to minimize the computer resources and to reduce the impact of H⁻ volume production on total extracted current to focus on the extraction of surface emitted H⁻ ions.

**Table 2.** Comparison of the H⁻ beam parameters for homogeneous and linear distributions of the H⁻ surface emission rate, the emission rate linear dependence is characterized by the values at 6 and 0 mm upstream the PE aperture.

| Emission rate | [Am$^{-2}$] | **20** | **2 - 37** | **37 - 2** |
|---|---|---|---|---|
| H⁻ total | [mA] | 21 | 25.5 | 13.6 |
| H⁻ volume | [mA] | 5.6 | 5.4 | 5.9 |
| H⁻ surface | [mA] | 15.5 | 20 | 7.6 |
| e | [mA] | 5 | 7 | 7 |
| e/H⁻ total | | 0.24 | 0.27 | 0.5 |

The result indicates that non-uniform emission rate impacts the total beam current and Cs-coverage on the PE-tip leads to higher H⁻ yield and lower e/H⁻. Approximately 80 % of the H⁻ extracted from the source is produced from the surface aperture close to PE tip. A fraction of H⁻ emitted from PE close to bulk plasma cannot be directly extracted. The fraction depends on the emission rate and the location of the meniscus.

## 4. Conclusion

A modified version of 3D-PIC MCC code ONIX with a single extraction aperture and non-periodic boundary conditions was implemented and used to model the beam formation region of the Linac4 ion source. The effect of the plasma parameters and surface production rate of negative ions from the PE on the extraction current have been illustrated in this work. The ONIX simulation results are in agreement with experimental measurements for a well caesiated source in terms of the extracted negative ions and co-extracted electron current.

ONIX simulations confirm that the extracted H⁻ beam is affected by the vertical magnetic filter field that induce horizontal asymmetry correlated to the co-extracted electrons. Simulations demonstrate that the asymmetry is larger for higher emission rate and lower e/H⁻. A ± 10% effect in the horizontal plane and ± 2% in the vertical plane is observed for 60 mA beam with e/H⁻ less than 1. The H⁻ ions generated from the surface induce a radial asymmetry and have an increased production near the PE-tips. Impact of the production rate and Cs-coverage gradient on the beam properties of the IS03 H⁻ source has been demonstrated.

## Acknowledgments

The authors thank the IPP Garching and LPGP-Orsay teams for the collaboration, their contributions and support during the development of the ONIX code. This work was supported by a grant from the Swiss National Supercomputing Centre (CSCS) under project ID s1061. We acknowledge PRACE for awarding access to the Fenix Infrastructure resources, which are partially funded from the European Union's Horizon 2020 research and innovation programme through the ICEI project under the grant agreement No. 800858.